\documentclass[sigplan,screen,nonacm]{acmart}

\AtBeginDocument{%
  }

\usepackage{microtype}
\usepackage{booktabs}
\usepackage{amsmath}
\usepackage{listings}
\usepackage{xcolor}
\usepackage{enumitem}

\usepackage{tikz}
\usetikzlibrary{arrows.meta,positioning,fit,calc}
\usepackage{tabularx}
\usepackage{booktabs}

\usepackage{graphicx}

\lstdefinestyle{agentos}{
  basicstyle=\ttfamily\small,
  columns=fullflexible,
  frame=single,
  breaklines=true,
  showstringspaces=false
}

\setcopyright{none}
\copyrightyear{2026}
\acmYear{2026}

\acmConference[AgenticOS '26]{The 1st Workshop on Operating Systems Design for AI Agents (AgenticOS), co-located with ASPLOS 2026}{March 23, 2026}{Pittsburgh, PA, USA}

\title[Grimlock]{Grimlock: Guarding High-Agency Systems with eBPF and Attested Channels}

\author{
Qiancheng Wu$^\ast$, Wenhui Zhang$^\ast$, Gan Fang, Sheng Mao, Biao Gao, David Levitsky, Shawna Murphy Butterworth, Rob Cameron \\
Roblox \\
\{qwu, wenhuizhang, gfang, smao, bgao, dlevitsky, sbutterworth, rcameron\}@roblox.com
}

\renewcommand{\shortauthors}{Qiancheng et al.}

\begin{abstract}
Agentic systems increasingly run user-authored orchestration code that invokes tools, spawns subtasks, and delegates work across machines and clouds. Although this high agency is productive, it creates a security problem: identity, authorization, provenance, and delegation are often pushed into application code, where they become difficult to enforce consistently and difficult to audit.

We present \emph{Grimlock}, an \emph{Agent Guard} that restores separation of concerns by moving trust enforcement into the sandbox substrate while leaving agent code unchanged. Grimlock uses \emph{eBPF-enforced traffic interception} to ensure that sandbox communication passes through a guard, and combines it with \emph{post-handshake attestation} bound to standard TLS~1.3 channel bindings. After a channel is established, the guard authorizes communication and mints short-lived, channel-bound \emph{scope tokens} that capture least-privilege delegation. At the receiving side, the destination guard re-validates identity, scope, and channel binding, terminates TLS, and releases plaintext to the destination sandbox only after policy checks succeed. kTLS provides an efficient dataplane for protected communication.

As a result, Grimlock offers a path toward transparent, auditable, and scope-bound agent-to-agent communication across heterogeneous multi-cloud environments, using commodity Linux primitives and without requiring changes to user-layer orchestration code.

\end{abstract}


\begin{document}

\maketitle
\let\thefootnote\relax\footnotetext{$^\ast$These authors contributed equally to this work.}

\setlength{\parskip}{0pt}
\setlength{\parindent}{0em}

\section{Introduction}

Agentic software increasingly acts as an orchestrator: it plans, calls tools, invokes other agents, and composes workflows across machines and clouds \cite{wu2024autogen,hong2024metagpt,liu2023agentbench,wang2025agents}. This yields a desirable property we call \textbf{high agency}: the user specifies intent without micromanaging infrastructure. But high agency often collapses security layering: application code begins to embed identity and authorization logic, duplicates policy checks, and stitches together ad-hoc credentials, reintroducing classic confused-deputy and privilege-management pitfalls \cite{hardy1988confused,Sandhu94accesscontrol}. The result is brittle trust, weak auditability, and poor portability.

We argue an Agentic OS should provide both high agency \emph{and} strict trust boundaries by separating concerns: user code handles orchestration, while a higher-privilege sandbox substrate enforces identity, authentication, authorization, and provenance. \textbf{Grimlock} realizes this split through eBPF no-bypass mediation and kTLS-bound post-handshake attestation for authenticated agent-to-agent communication.

\textbf{Why eBPF.}
High-agency agents are untrusted and highly dynamic: they may spawn processes, open arbitrary sockets, and call tools that live outside the application address space. Purely library- or runtime-level enforcement is therefore easy to bypass.
Grimlock uses eBPF because it enables \emph{OS-enforced, application-transparent} mediation at the sandbox boundary. eBPF can interpose on ingress/egress, associate flows with stable sandbox identity, and \emph{force} all traffic through a controlled path such as redirecting sockets to a guard proxy) without modifying agent code. 


\textbf{Why post-handshake attestation.}
Attested TLS systems can be categorized by when attestation evidence is generated and bound relative to connection establishment \cite{SardarSeatIntraVsPost}. We adopt post-handshake attestation because it preserves interoperability with commodity TLS~1.3 stacks (no handshake modifications) while still binding evidence to the established channel using exporter-based channel bindings and freshness nonces. This designs helps mitigate replay, diversion, and relay attacks.
\cite{SardarSeatIntraVsPost,RFC9261,RFC9266}. 

We present \textbf{Grimlock}, an Agentic OS guard layer that secures agent communication by enforcing a guarded, no-bypass datapath and channel-bound authorization.

Contributions of this paper include:
\begin{itemize}[leftmargin=*]
  \item \textbf{Transparent mediation:} an architecture for transparently intercepting and routing sandbox communication through a guard layer.
  \item \textbf{No-bypass enforcement:} an eBPF-based design for mandatory mediation at the sandbox boundary, preventing agent traffic from evading the guard.
  \item \textbf{kTLS dataplane:} a kTLS-based TLS~1.3 datapath that keeps record processing in kernel space while remaining transparent to agent code.
  \item \textbf{Scope-bound A2A authorization:} a post-handshake, channel-bound attestation and authorization design for secure agent-to-agent communication.
\end{itemize}

\section{Design of Grimlock}
\begin{figure}[t]
  \centering
  \includegraphics[width=0.95\linewidth]{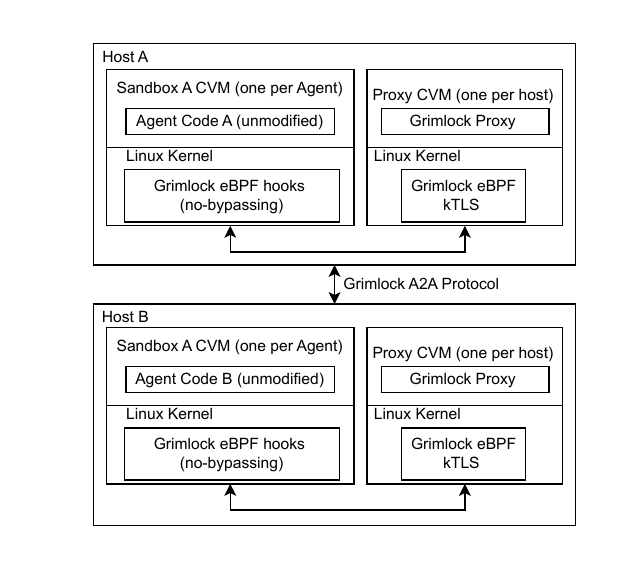}
  \caption{Grimlock deployment. Each host runs a per-agent sandbox Confidential VM (CVM) and a per-host proxy CVM. eBPF enforces no-bypass mediation and steers traffic through the proxy. Proxies communicate via the Grimlock A2A Protocol over TLS~1.3 (kTLS-assisted datapath).}
  \label{fig:grimlock-a2a}
\end{figure}

Grimlock provides substrate-level security for high-agency agents by enforcing a no-bypass network mediation point at the sandbox boundary and binding authorization to active encrypted channels, shown in Figure.~\ref{fig:grimlock-a2a}.

\setlength{\abovecaptionskip}{0pt}   
\setlength{\belowcaptionskip}{0pt}   

\subsection{Threat Model}

We assume an active network adversary capable of eavesdropping, replay, relay/diversion, and man-in-the-middle attacks. We also consider a malicious or buggy agent runtime that may attempt to bypass mediation (e.g., direct sockets, alternate stacks) or escalate scope. 

We do \emph{not} trust user-layer code to correctly implement identity or authorization, and we treat the host network and surrounding infrastructure as untrusted. 

Grimlock targets three properties: (i) \textbf{no-bypass}, all sandbox traffic must traverse the guard; (ii) \textbf{channel binding}, authorization artifacts are bound to a specific established channel; and (iii) \textbf{least privilege}, delegation propagates auditable, scoped permissions.

\subsection{Architecture and Data Path}
Grimlock comprises per-agent sandboxes (CVMs) and a per-host guard proxy. Each agent executes inside a confidential, attestable CVM. A CVM is trusted only after remote attestation succeeds under operator policy, covering the boot chain and measured guest software stack (guard and kTLS kernel mechanisms).

Grimlock enforces mandatory mediation at the sandbox boundary. eBPF hooks interpose on all sandbox ingress/egress and redirect all traffic to the local guard proxy CVM, preventing agents from bypassing policy. The proxy establishes a standard TLS~1.3 handshake (using kTLS for efficient record processing) and then runs post-handshake attestation bound to the established channel via exporter-derived channel bindings. After successful appraisal by a verifier/issuer, a short-lived, channel-bound \emph{Scope Token} encoding least-privilege delegation is minted. Traffic is then forwarded to the destination host, where the receiving guard re-validates the attestation result, token scope, audience, expiry, and channel binding; terminates TLS; and releases plaintext to the destination sandbox only after policy checks succeed.

\subsection{Grimlock A2A Protocol and Scope Token}
Agent A initiates A2A communication by issuing a standard socket \texttt{connect()} from within its sandbox. Grimlock enforces \emph{no-bypass} at the sandbox boundary using eBPF-based mediation. The guard maintains per-flow state (flow metadata $\rightarrow$ source sandbox identity, source and destination ip, destination port, requested-scope hash, expiry) and coordinates authorization over a long-lived host-to-host control channel. This amortizes setup while allowing connection establishment and guard authentication and authorization to proceed concurrently.

Between guards, Grimlock establishes a standard TLS~1.3 channel. The first contact with a peer performs full mutual authentication and may cache a peer authentication context for reuse on subsequent connections. Once the TLS session is established, kTLS is configured on the guard-to-guard channel so that record encryption and decryption occur in the kernel dataplane without requiring any changes to agent code.

Authorization is enforced as a \emph{post-handshake} gate bound to the established channel. Immediately after the TLS handshake completes, and before any application payload is released, the guards compute a channel binding:

$\mathsf{cb}=\mathsf{Exporter}(\mathsf{TLS},\mathsf{label},\mathsf{ctx})$

where $\mathsf{ctx}$ includes a fresh nonce, the intended audience, and the requested delegation scope. The responder returns TEE evidence that commits to $H(\mathsf{cb})$. A verifier appraises this evidence under operator policy and mints a short-lived, channel-bound \emph{Scope Token}. The receiving guard validates the token and channel binding before terminating TLS and releasing plaintext into the destination sandbox.

\section{Conclusion}

High-agency agentic software demands both flexibility and strong trust boundaries. Grimlock delivers this separation by enforcing no-bypass traffic mediation with eBPF and securing agent-to-agent communication over TLS~1.3, with kTLS providing an efficient protected dataplane.

\begin{acks}
We thank the AgenticOS'26 reviewers and organizers for feedback.
\end{acks}

\bibliographystyle{ACM-Reference-Format}
\bibliography{tee}

\end{document}